\documentclass[prd,onecolumn,floatfix,letterpaper,amsmath,amssymb,nofootinbib,superscriptaddress]{revtex4}

\usepackage{graphicx}
\usepackage{amsfonts}
\usepackage{color}
\usepackage{bm}
\usepackage{mathrsfs}
\usepackage{epstopdf}
\usepackage{url}
\usepackage[footnotesize]{caption}
\usepackage{footnote}
\usepackage[caption=false]{subfig}
\usepackage{textcomp}

\usepackage{amsmath, amssymb, graphicx, float}

\usepackage[final]{feynmp}
\usepackage{verbatim} % for muitiple line comment
\makeatletter
\def\endfmffile{%
	\fmfcmd{\p@rcent\space the end.^^J%
			end.^^J%
			endinput;}%
	\if@fmfio
		\immediate\closeout\@outfmf
	\fi
	\ifnum\pdfshellescape=\@ne
		\immediate\write18{mpost \thefmffile}%
	\fi}

\newcommand{\mathsym}[1]{{}}

\clearpage
\cleardoublepage
\newcommand{\red}[1]{{\textcolor{red}{#1}}}

\begin{document}
%\begin{fmffile}{G4}

\setlength{\unitlength}{1mm}

\title{The boundary effect of anomaly-induced action}

   \author{Che-Min Shen}
   	\email{f97222024@ntu.edu.tw}
    \address{Department of Physics, National Taiwan University, Taipei 10617, Taiwan, R.O.C.}
    \address{Leung Center for Cosmology and Particle Astrophysics, National Taiwan University, Taipei 10617, Taiwan, R.O.C.}

       \author{Keisuke Izumi}
   	\email{izumi@phys.ntu.edu.tw}
    \address{Leung Center for Cosmology and Particle Astrophysics, National Taiwan University, Taipei 10617, Taiwan, R.O.C.}

    \author{Pisin Chen}
   	\email{chen@slac.stanford.edu}
\address{Department of Physics, National Taiwan University, Taipei 10617, Taiwan, R.O.C.}
\address{Leung Center for Cosmology and Particle Astrophysics, National Taiwan University, Taipei 10617, Taiwan, R.O.C.}
\address{Graduate Institute of Astrophysics, National Taiwan University, Taipei, Taiwan 10617}
\address{Kavli Institute for Particle Astrophysics and Cosmology, SLAC National Accelerator Laboratory, Stanford University, Stanford, CA 94305, U.S.A.}

\begin{abstract}
We discuss the boundary effect of anomaly-induced action in two-dimensional spacetime, which is ignored in previous studies.
Anomaly-induced action, which gives the stress tensor with the same trace as the trace anomaly, can be represented in terms of local operators by introducing an auxiliary scalar field.
Although the degrees of freedom of the auxiliary field can in principle describe the quantum states of the original field, the principal relation between them was unclear.
We show here that, by considering the boundary effect, the solutions of classical auxiliary fields are naturally related to the quantum states of the original field.
We demonstrate this conclusion via several examples such as the flat, black hole and the de Sitter spacetime.
\end{abstract}

\maketitle
\section{Introduction}
In the absence of a well-developed theory of quantum gravity, the semiclassical approach, that is  quantum field theory in (classical) curved spacetimes, has been applied widely to study quantum corrections to general relativity~\cite{BD}.
The semiclassical approach, where the quantum divergences of fields are covariantly renormalized, gives the (one-loop) effective action.
The expectation value of the stress tensor of quantum matter fields can be also derived with this procedure.
The result suggests that, even in conformal field theories, a nonzero trace of stress tensor arises by the renormalization.
This nonzero trace of stress tensor is called the trace (or conformal) anomaly~\cite{Deser:1976yx,Duff:1977ay,BD}.
% which is a purely quantum effect.

In principle, we can obtain the expectation value of the stress tensor of quantum matter fields %from the (one-loop) effective action
in this semiclassical approach (i.e. the quantum field theory in curved spacetimes).
Meanwhile, we have a practical problem; the calculation is so complicated that
there is no explicit expression of the effective stress tensor in general background spacetimes.
We need to derive the effective stress tensor individually in each spacetime that we are interested in.
Because of the complicated calculations, we usually rely on, for instance, numerical and/or approximation approaches, even in simple common spacetimes such as Schwarzschild spacetime~\cite{BHcalcalation}.
One way to tackle with this problem is rebuilding the corresponding \emph{anomaly-induced action}~\cite{Polyakov:1981rd,Fradkin:1983tg,Riegert:1984kt}.
% by functional integration.
Although the anomaly-induced action is not always equal to the (one-loop) effective action from the original semiclassical approach, % and it might differ from the original one by some conformal invariant terms,
it can be expected and has been checked in some specific cases that in two-dimensional spacetime the anomaly-induced action can exactly describe %the original result
the stress tensor of quantum field in vacuum state~\cite{Balbinot:1999vg}.
In four-dimensional spacetime, the anomaly-induced action could not correctly reproduce the original semiclassical result, but we would still be able to get at least some feeling.
This approach has been applied widely to study the quantum stress tensor in curved spacetimes~\cite{Mottola:2006ew,Anderson:2007eu}, black-hole physics ~\cite{Balbinot:1999vg,Anderson:2007eu,Mukhanov:1994ax} and cosmology~\cite{Fabris:1998vq,Fabris:2000gz,Pelinson:2002ef,Bilic:2007gr}.
The anomaly-induced action is naturally built in non-local form, and can be localized by introducing an additional auxiliary\footnote{
The word, ``auxiliary", often be used for fields without physical dynamics. Meanwhile, the auxiliary field here obeys an equation of motion, and thus contains a dynamical degree of freedom.
}
scalar field~\cite{Mazur:2001aa}.
Different solutions of the auxiliary scalar fields could describe the effects of different quantum states of the original conformal matter field.
Although there are attempts to find the correspondence between the quantum states of the original field and the solutions of the auxiliary scalar field,
so far we have not known the general principle behind it.

%Although we expect that imposing boundary conditions on those fields is equivalent to selecting quantum states, we in fact lose reference to particular quantum (vacuum) state.

%As we just mentioned, imposing boundary conditions on those auxiliary fields is equivalent to choosing quantum states (of quantum matter fields) in the original semiclassical approach and describe the macroscopic effect of quantum matter.
%Since the boundary terms can fix a part of dynamics, we expect that they could play some important roles which has always been neglected in the previous works. This assumption might be reasonable in Minkowski spacetime or other asymptotic-flat spacetime without horizon (boundary), because we can expect the boundary terms would decay fast enough when approaching the infinite. However, when facing the spacetime which is not asymptotic-flat or with a horizon (boundary), the assumption cannot be expected and will fail in general. Therefore motivated by some possible existing boundary effect of the semiclssical ("anomaly") approach, our main purpose in this paper is to restore the boundary terms which is always neglected during the derivation of anomaly induced action and see what is the role of this effect.

In this paper, we take into consideration the boundary effect in the discussion of anomaly-induced action,
which has been neglected in the previous works~\cite{Polyakov:1981rd,Fradkin:1983tg,Riegert:1984kt,Balbinot:1999vg,Mottola:2006ew,Anderson:2007eu}.
This effect can be important, for instance, if we consider a black hole spacetime.
We sometimes construct a quantum field theory only in the outside of the horizon.
Then, the horizon is the boundary of the spacetime where the quantum field theory is defined.
Moreover, the boundary effect is important even at spatial infinities.
In the construction of quantum field theories in a spacetime with spatial infinities, we first construct the quantum field theory in a finite region, and then take the limit where the boundary goes to the spatial infinity.
In this procedure, the boundary action manifestly affects the result.

After deriving the generic form of the anomaly-induced action with the boundary effect in two-dimensional spacetime,
we apply the result to simple cases; the flat spacetime, two-dimensional Schwarzschild black hole (a black-hole-like spacetime which is corresponding to the time and radial parts of the four-dimensional Schwarzschild black hole) and de Sitter spacetime.
We find a natural relation between the quantum states of the original field and the
solutions of the auxiliary scalar field, after taking the appropriate limit of the boundaries.
For instance, in the flat space, taking the limit where the boundary tangent to Rindler time goes to Rindler horizon, we have the stress tensor of the Rindler vacuum state.
In the similar analysis with the two-dimensional Schwarzschild metric, we can naturally obtain the stress tensors of the Boulware, Hartle-Hawking and Unruh states.
In de Sitter spacetime, the stress tensors of the Bunch-Davies state and the vacuum stress tensor in static coordinate are straightforwardly derived.

The structure of this paper is organized as follows. In Sec.~\ref{review}, we give a short review of the trace (conformal) anomaly and the anomaly-induced action (Polyakov action) in two dimensional spacetime.
In Sec.~\ref{WB},
%in order to take boundary effect which is always neglected in previous works into consideration, during the derivation of anomaly induced action,
we introduce the boundary term. % responding to the original counter terms responsible for usual renormalization of quantum fields in curved spacetime (in 2-dim).
%We thus show that the anomaly induced action should be modified due to boundary effect.
%It turns out that the additional scalar field $\varphi$ should satisfy not only the original (2nd-order differential equation) equation of motion but also additional constraints on boundary.
The boundary terms give the boundary conditions for the auxiliary scalar field, which constrain on the solutions of the auxiliary scalar field.
%As a result, the corresponding stress tensor have less degree of freedom remained in the general solution than what people obtain before.
In Sec.~\ref{EX}, as examples, we apply our result to several common spacetimes.
%We examine the general solution of the auxiliary scalar field and the corresponding stress tensor for all of it.
%As a result, the additional boundary constraints reduce the degree of freedom of the general solution and interestingly it also provide us a criteria to choosing the correct boundary conditions of auxiliary field for particular quantum states.
We see that the boundary conditions lead to the stress tensor of the naturally corresponding state.
Finally, we give a summary and discussion in Sec.~\ref{Sum}.

\section{Review of anomaly-induced action}\label{review}

On a curved spacetime, the conformal anomaly appears through the renormalization of the stress tensor.
The expectation value of the stress tensor diverges even for the linear field theory\footnote{
This divergence appears even in the flat spacetime, which is the vacuum energy.
In a theory without gravity we can just ignore it, because it is coupled only with gravity.
In a gravitational theory, however, since it can be the source of gravity, we need to renormalize it.
Usually, we assume that the renormalized vacuum energy is tiny, which might explain the acceleration of the Universe.
Nevertheless, there is no natural reason for its smallness, which is the well-known cosmological constant problem.
This issue is beyond the scope of
this paper and we will not dwell on it further.
}
and the renormalization is required.
The counter terms represented by in geometric forms are introduced for the renormalization
and, in even-dimansional spacetime, the anomalous contribution appears in the gravitational equation.
This contribution violates the conformal symmetry even if the original action for the fields possesses the symmetry, and thus it is called the conformal (or trace) anomaly~\cite{Deser:1976yx,Duff:1977ay,BD}.
%It is hard to write down the calssical local action for the anomalous contribution straightforwardly.
The action rebuilt from this anomalous contribution is written by the nonlocal and geometric functions.
Meanwhile, by introducing scalar fields, the classical action for the anomalous terms can be expressed in a local form.
The local form is useful for applicative discussions, such as cosmology.
In %the beginning of
this section, we briefly review the idea of the effective local action for the trace anomaly~\cite{Mazur:2001aa,Mottola:2006ew}.

We first consider a Lagrangian of a scalar field in $n$-dimensinal spacetime:
\begin{eqnarray}
{\mathcal{L}_{cl}} = \frac{1}{2}{\nabla ^\mu }\phi {\nabla _\mu }\phi  - \frac{1}{2} (m^2 + \xi R) {\phi ^2},\label{OLag}
\end{eqnarray}
where $m$ is the mass of the scalar field and $\xi$ is a dimensionless constant.
% $\xi (n) := \frac{{(n - 2)}}{{4(n - 1)}}$.
The effective Lagrangian including the one-loop contributions for this scalar field can be derived as~\cite{BD}
\begin{eqnarray}
{L_{eff}} = \frac{i}{2}\mathop {\lim }\limits_{x \to x'} \int_{{m^2}}^\infty  {d{m'^2}} G_F^{DS}(x,x';m'^2), \label{Lan}
\end{eqnarray}
where $G_F^{DS}$ is the DeWitt-Schwinger representation of the Feynman propagator $G_F$.
By the DeWitt-Schwinger expansion, this effective Lagrangian can be expanded as
\begin{eqnarray}
L_{eff}(x) \approx \frac{1}{2(4\pi)^{n/2}} \sum\limits_{j=0}^{\infty} a_{j}(x) m^{n-2j} \Gamma(j-\frac{n}{2}),
\end{eqnarray}
where $a_j$'s are written in the geometric forms:
\begin{eqnarray}
&&a_{0}(x) = 1, \\
&&a_{1}(x) = (\frac{1}{6}-\xi)R, \\
&&a_{2}(x) = \frac{1}{2}(\frac{1}{6}-\xi)^2R^2 + \frac{1}{180}(R_{\mu\nu\alpha\beta}R^{\mu\nu\alpha\beta}-R_{\mu\nu}R^{\mu\nu}) -\frac{1}{6}(\frac{1}{5}-\xi)\Box R, \\
&& \ldots  \ . \nonumber
\end{eqnarray}
The gamma function $\Gamma(j-n/2)$ diverges when its argument is naught or a negative integer,
and thus we must introduce counter terms to renormalize these divergent parts.

Since these divergences mostly stem from the curvature effect of spacetime,
we expect that the counter terms are written with the geometric forms.
Indeed,  we can introduce the corresponding counter terms to cancel all divergent parts of the effective Lagrangian.
The renormalized effective Lagrangian is defined as
\begin{eqnarray}
L_{ren} := L_{eff} - L_{ct},
\end{eqnarray}
where $L_{ct}$ is the Lagrangian of the counter terms.
The renormalized stress tensor is derived to be
\begin{eqnarray}
\left\langle T_{\mu\nu}\right\rangle_{ren} := \frac{-2}{\sqrt{-g}} \frac{\delta S_{ren}}{\delta g^{\mu\nu}} = \frac{-2}{\sqrt{-g}} \frac{\delta \int {{d^n}x\sqrt { - g} } {L_{ren}}}{\delta g^{\mu\nu}}.
\end{eqnarray}
In order to take advantage of conformal symmetry, hearinafter we will consider a conformally coupled scalar field, i.e. the case with $\xi (n) := \frac{{(n - 2)}}{{4(n - 1)}}$, and $m = 0$.
Due to the conformal symmetry, the classical stress tensor is traceless.
In even-dimension, however, the renormalized stress tensor for the conformal scalar field has a nonzero trace
\begin{eqnarray}
g^{\mu\nu} \left\langle T_{\mu\nu}\right\rangle_{ren} =  - \frac{a_k(x)}{(4\pi)^k} \qquad \mbox{with} \qquad k=\frac{n}{2}, \qquad k\in\mathbb{N},
\end{eqnarray}
originating from the counter terms.
This term manifestly violates the conformal symmetry, and thus it is called the trace (or conformal) anomaly.

Hereinafter, we investigate the case in 2-dimensional spacetime as the simplest example.
We start with the derivation of the non-local action for this anomalous contribution.
The Wess-Zumino (WZ) action is useful for this derivation, which is defined as
\begin{eqnarray}
\Gamma_{W\!\!Z}[\bar{g},\sigma] := S[\bar{g}] - S[g] ,
\end{eqnarray}
with
\begin{eqnarray}
\bar{g}_{\mu\nu}:=\exp(-2\sigma) g_{\mu\nu}. \label{CT}
\end{eqnarray}
Due to the conformal symmetry, before introducing counter terms, the action is conformally invariant, i.e. we have $S_{eff}[\bar{g}] = S_{eff}[g]$.
This makes the relation of the renormalized WZ action to that for the counter terms as
\begin{eqnarray}
\Gamma_{W\!\!Z}[\bar{g},\sigma] &=& S_{ren}[\bar{g}] - S_{ren}[g] \nonumber\\
&=& \bigl[S_{eff}[\bar{g}] - S_{eff}[g]\bigl] - \bigl[S_{ct}[\bar{g}] - S_{ct}[g]\bigl] \nonumber\\
&=& 0 - \bigl[S_{ct}[\bar{g}] - S_{ct}[g]\bigl].  \label{Gref=Gdiv}
\end{eqnarray}
From the WZ action we could read the form of the renormalized action $S_{ren}$.
However, the renormalized action derived from the WZ action has ambiguity; adding conformally invariant terms $S_{conf}$ to the obtained action $S_{ren}$,
the new action $S_{ren}+S_{conf}$ still gives the same WZ action.
%That is, a general form of the renormalized action is $S_{ren}=S_{anom}+S_{conf}$.
This ambiguity is supposed to be partially related to the degrees of freedom of the quantum state.
Meanwhile, all information of the trace anomaly is definitely included in $S_{ren}$, and thus  the renormalized action that we can read from the WZ action is called the anomaly-induced action $S_{anom}$, i.e.
% is definitely included in $S_{anom}$.
%Therefore, $S_{anom}$ is called the anomaly action, and it satisfy the following relation:
\begin{eqnarray}
\Gamma_{W\!\!Z}[\bar{g},\sigma] = S_{anom}[\bar{g}] - S_{anom}[g]. \label{GWZ=Sanom}
\end{eqnarray}

In the case of a conformally coupled scalar field in two-dimensional spacetime, the divergent parts of the effective Lagrangian (\ref{Lan}) is written as
\begin{eqnarray}
S_{ct} = \frac{-1}{24\pi} \mathop {\lim }\limits_{n \to 2}  \int d^2x \sqrt{-g} \frac{ R}{ (n - 2)}.
\label{Ldiv2}
\end{eqnarray}
Substituting this into eq.(\ref{Gref=Gdiv}), we can derive the WZ action as
\begin{eqnarray}
\Gamma_{W\!\!Z}[\bar{g},\sigma] &=&
  \frac{1}{24\pi} \lim \limits_{n \rightarrow 2}[\frac{  \int d^2x \sqrt{-\bar{g}} \bar{R} - \int d^2x \sqrt{-g} R  }{n-2}] \nonumber \\
&=& - \frac{1}{24\pi}  \int d^2x \sqrt{-\bar{g}} [\sigma \bar{R} - \sigma \bar{\Box} \sigma ] \nonumber\\
&=& \frac{1}{96\pi} \int d^2x \int d^2x' \sqrt{-\bar g} \sqrt{-\bar g'} \bar R(x) \bar D_2(x,x') \bar R(x') \nonumber\\
&&\qquad\qquad
-\frac{1}{96\pi} \int d^2x \int d^2x' \sqrt{-g} \sqrt{-g'} R(x) D_2(x,x') R(x') .
\label{Gren1}
\end{eqnarray}
From eq.(\ref{GWZ=Sanom}) and eq.(\ref{Gren1}), we can find
%Take advantage of conformal symmetry of this theory, we can obtain the corresponding non-local anomalous action:
\begin{eqnarray}
S_{anom}[g] = \frac{1}{96\pi} \int d^2x \int d^2x' \sqrt{-g} \sqrt{-g'} R(x) D_2(x,x') R(x') ,
\label{Sanom}
\end{eqnarray}
where $D_2$ is the inverse operator of D'Alembert operator, i.e.
\begin{eqnarray}
\Box D_2(x,x') = - \frac{\delta(x-x')}{\sqrt{-g}}.
\end{eqnarray}
In the last equality of eq.(\ref{Gren1}), we have imposed the symmetric condition of $D_2$, i.e. $D_2(x,x')=D_2(x',x)$, and used the relation
\begin{eqnarray}
2\sqrt{- \bar g}\bar \Box \sigma= \sqrt{- \bar g} \bar R - \sqrt{-g} R, \label{s=R}
\end{eqnarray}
which is obtained from the conformal transformation (\ref{CT}).

This non-local anomaly-induced action can be localized by introducing a real scalar field $\varphi$ which is defined as
\begin{eqnarray}
\varphi(x) := \int d^2x' D_2(x,x') R(x'). \label{phiD2R}
\end{eqnarray}
Operating this by the D'Alembert operator, we can obtain\footnote{
Although the definition (\ref{phiD2R}) seems equivalent to (\ref{boxphi}),
for derivation of eq.(\ref{phiD2R}) from eq.(\ref{boxphi}) we need the double integrations.
Therefore eq.(\ref{phiD2R}) has information of eq.(\ref{boxphi}) and two integration constants, i.e. using a specific inverse function $D_2$ is indeed equivalent to choosing a specific particular solution for eq.(\ref{boxphi}) here.
Meanwhile, in four-dimensional spacetime, in order to derive the localized anomaly action, we need to introduce two scalar fields~\cite{Shapiro:1994ww,Mazur:2001aa}. They should share the same green function (inverse operator) analogous to $D_2$.
Thus, the differential equations that two scalars should satisfy are no longer independent.
}
\begin{eqnarray}
\Box \varphi = -R.\label{boxphi}
\end{eqnarray}
The degrees of freedom of the integration constants are absorbed into those of the Green function $D_2(x,x')$.
Equation (\ref{boxphi}) can be obtained from the following action
%Now the localized version of the anomalous action (\ref{Sanom}) can be expressed in terms of the auxiliary scalar field $\varphi$ as
\begin{eqnarray}
%\Rightarrow
S_{anom}[g,\varphi] = \frac{1}{96\pi} \int d^2x \sqrt{-g} [ g^{\mu\nu} \nabla_{\mu}\varphi \nabla_{\nu}\varphi -2\varphi R],
\label{SanomLocal}
\end{eqnarray}
We can check that this action is reduced to the anomaly action (\ref{Sanom}) after substituting eq.(\ref{phiD2R}).
%Thus, this is the local anomalous action, which
This gives the same dynamics as the non-local action (\ref{Sanom}).
The corresponding stress tensor can be obtained by the variation with respect to the metric $g_{\mu\nu}$, and its explicit form is
\begin{eqnarray}
{\rm{T}}_{\mu \nu }^{anom}
&:=&  - \frac{2}{{\sqrt { - g} }}\frac{{\delta {S_{anom}}}}{{\delta {g^{\mu \nu }}}} \\
&=& \frac{1}{{24\pi }}[({R_{\mu \nu }} - \frac{1}{2}R{g_{\mu \nu }})\varphi  - {\nabla _\mu }{\nabla _\nu }\varphi  + {g_{\mu \nu }}\Box\varphi  - \frac{1}{2}({\nabla _\mu }\varphi )({\nabla _\nu }\varphi ) + \frac{1}{4}{g_{\mu \nu }}({\nabla _\alpha }\varphi )({\nabla ^\alpha }\varphi )].
\label{Tanom}
\end{eqnarray}
This trace consists with the well-known trace anomaly,
\begin{eqnarray}
{g^{\mu \nu }}{\rm{T}}_{\mu \nu }^{anom} = \frac{1}{{24\pi }}\Box\varphi  =  - \frac{1}{{24\pi }}R.
\end{eqnarray}
Therefore, it is concluded that scalar field action (\ref{SanomLocal}) describes the anomalous contribution.

\section{anomaly-induced action with boundaries}\label{WB}
%Usually we ignore all surface terms in the derivation of this theory(approach). When considering asymptotic-flat spacetime, this assumption might make sense because we expect those surface terms will vanish far away(in the infinite boundary).
%However, once the spacetime we are interested in is no longer asymptotic-flat or with a boundary, for example, de-Sitter space or Schwarzschild black hole, we will expect the surface terms no longer vanish in general and thus we should take care of them carefully during the derivation of the anomaly action.

In this section, we introduce the surface terms (i.e. the boundary effect), which was ignored in the previous works~\cite{Polyakov:1981rd,Fradkin:1983tg,Riegert:1984kt,Balbinot:1999vg,Mottola:2006ew,Anderson:2007eu}.
The boundary effect is important not only for bounded spacetimes but also for unbounded ones.
Considering a spacetime with a horizon, the surface term fixes the boundary condition on the horizon.
Meanwhile, at the (spatial) infinity, the surface term constrains on the asymptotic behavior.

We indeed need to take only timelike boundaries into consideration.
In the standard way to derive classical equation of motion, we take the variation of the action while fixing the initial and final states.
Even if the surface terms on the spacelike boundaries (i.e. the initial and final hypersurfaces) are introduced, the final form of the stress tensor derived by the variation of the action is not affected.
We thus ignore the contribution of spacelike boundaries. % when we write down the anomaly action.

%In order to consider (this theory) again in a bounded region, we need to modify the divergent part of effective action with a corresponding boundary term. In 2-dim case, the divergent part is indeed proportion to Ricci scalar, so the corresponding boundary term should also proportion to the well-known Gibbons-Hawking term.
%Therefore in the following, we will start with (the modified divergent counter action).
%In a bounded region, the counter term should accompany with the surface term.
In two-dimensional case, since the counter term is proportional to the Einstein-Hilbert term, the corresponding boundary term is the Gibbons-Hawking term~\cite{Gibbons:1976ue,York:1972sj}.
The action of the counter term including boundary term is
\begin{eqnarray}
S_{ct}[g] := \frac{-1}{24\pi} \frac{\int_{\mathcal{M}} d^2x \sqrt{-g} R + 2\int_{\Sigma} d^1x \sqrt{-\gamma} K }{n-2},
\end{eqnarray}
where $\mathcal{M}$ is two-dimensional spacetime and $\Sigma$ is the timelike boundary.
For convenience in the later discussion, we introduce an arbitrary scalar function $f(x)$ which is unity on the boundary and arbitrary elsewhere, i.e.
\begin{eqnarray}
f(x) = 1, \qquad x \in  \Sigma.
\end{eqnarray}
Using this scalar function, we rewrite the action of the counter terms in
\begin{align}
S_{ct}[g] %:= \frac{1}{24\pi} \frac{\int d^2x \sqrt{-g} R + 2\int d^1x \sqrt{-\gamma} K }{n-2}
= \frac{-1}{24\pi} \frac{\int d^2x \sqrt{-g} R + 2\int d^1x \sqrt{-\gamma} f K }{n-2}.
\end{align}

As is the case in eq.(\ref{Gref=Gdiv}), the corresponding WZ action is transformed into
\begin{align}
\Gamma_{W\!\!Z}[\bar{g},\sigma]
&= \frac{1}{24\pi} \lim \limits_{n \rightarrow 2}[\frac{ ( \int d^2x \sqrt{-\bar{g}} \bar{R} + 2\int d^1x \sqrt{-\bar{\gamma}} \bar{K} ) - (\int d^2x \sqrt{-g} R + 2\int d^1x \sqrt{-\gamma} K ) }{n-2}] \nonumber\\
&= - \frac{1}{24\pi} \{ \int d^2x \sqrt{-\bar{g}} [\sigma \bar{R} - \sigma \bar{\Box} \sigma ] + \int d^1x (\sqrt{-\gamma} \sigma K + \sqrt{-\bar{\gamma}} \sigma  \bar{K} ) \} \nonumber\\
&=: S_{anom}[\bar{g}] - S_{anom}[g].
\label{WZb}
\end{align}
%Similar to last section, by using the following relation
We have the relation analogous to eq.(\ref{s=R})
\begin{align}
2\sqrt{-\bar{g}} \bar{L}_f \sigma %= 2\sqrt{-\bar{g}}  [ - \bar{\Box}\sigma + \bar{\nabla}_{\mu} f \bar{n}^{\mu} \bar{n}^a \bar{\nabla}_{\alpha} \sigma]
= \sqrt{-g} [ R + 2\nabla^{\mu}(n_{\mu}fK) ] - \sqrt{-\bar{g}}[\bar{R} + 2 \bar{\nabla}_{\mu} ( \bar{n}^{\mu} f\bar{K} ) ]. \label{s=RK}
\end{align}
Here, $n^{\mu}$ is the unit normal vector on boundary and does not need to be fixed elsewhere.
$L_f$ is an operator defined as
\begin{eqnarray}
L_{f} := ( -\Box + \nabla^{\mu}f n_{\mu}n_{\nu}\nabla^{\nu} ).
\end{eqnarray}
The operator $L_f$ satisfies the following relation with arbitrary functions $h_1$ and $h_2$
\begin{eqnarray}
\int d^2x \sqrt{-g} h_1(x) L_f h_2(x) = \int d^2x \sqrt{-g} h_2(x) L_f h_1(x),
\end{eqnarray}
and thus $L_f$ is a self-adjoint operator.
%and after some straightforward but tedious calculation,
Using the relation (\ref{s=RK}), we can read the non-local anomalous action from eq.(\ref{WZb})
\begin{align}
S_{anom}[g] = \frac{1}{96\pi} \Biggl[ \int d^2x \int d^2x' \sqrt{-g} \sqrt{-g'} ( R(x) + 2\nabla^{\mu}(n_{\mu}fK) ) D_f(x,x') ( R(x') + 2\nabla'^{\mu}(n'_{\mu} f' K') ) \qquad&&
\nonumber\\
- 4 \int d^2x \sqrt{-g} f K^2 \Biggr].&&
\label{SanomB}
\end{align}
Here, $D_f$ is the symmetric inverse operator of $L_f$, which is defined by
\begin{eqnarray}
L_f D_f(x,x') = - \frac{\delta(x-x')}{\sqrt{-g}}, \qquad D_f(x,x')=D_f(x',x).
\end{eqnarray}

As the derivation of the local anomaly-induced action in the previous section, we introduce a real scalar field $\varphi$. % to localize the anomalous action.
The scalar field $\varphi$ is defined by
%This non-local action can be localized by introducing an auxiliary scalar field $\varphi$ which is defined by
\begin{eqnarray}
\varphi := \int d^2x' \sqrt{-g} D_f(x,x') [ R' + 2\nabla^{'}_{\mu}(n'^{\mu}f'K') ].
\end{eqnarray}
Operating $L_f$ to this equation, we have
\begin{eqnarray}
L_f \varphi = R + 2\nabla^{\mu}(n_{\mu}fK).
\label{EOMofPhi}
\end{eqnarray}
Considering the following action
%Again the localized version of the anomaly action (\ref{SanomB}) can be expressed in terms of the auxiliary scalar field $\varphi$ as
\begin{eqnarray}
S_{anom}[g] &=& \frac{1}{96\pi} \{ \int d^2x \sqrt{-g} [ \varphi L_f \varphi - 2 \varphi ( R(x) + 2\nabla^{\mu}(n_{\mu}fK) ) ] - 4 \int d^2x \sqrt{-g}f K^2 \} \\
&=& \frac{1}{96\pi} \{ \int {d^2}x\sqrt { - g} ( - \varphi \Box \varphi  - 2\varphi R) + \int {{d^2}x\sqrt { - g} f [ ({n_\mu }{\nabla ^\mu}\varphi) ( -{n_\nu }{\nabla ^\nu }\varphi  + 4K) - 4K^2]} \nonumber \\
&& + \int {{d^1}x\sqrt \gamma  (\varphi {n_\mu }{\nabla ^\mu }\varphi  - 4\varphi K)}  \},
\label{anb}
\end{eqnarray}
this action gives eq.~(\ref{EOMofPhi}) and, substituting eq.~(\ref{EOMofPhi}), this action is reduced into the non-local action (\ref{SanomB}).
Therefore, this is the localized anomaly action that we want.

Now we choose the useful form of the scalar function $f$.
Because $f$ is an arbitrary function except that it should be unity on the boundary, we can  consider the following $f$ function:
\begin{eqnarray}
f_\delta(\lambda) :=
\left\{
\begin{array}{cc}
\frac{1}{2}[\cos(\frac{\lambda\pi}{\delta})+1], & \ (0<\lambda\leq \delta) \\
0, &\ (\lambda\geq \delta) \\
\end{array}
\right.
\end{eqnarray}
where $\lambda$ is the affine parameter\footnote{We set the affine parameter $\lambda$ to be zero on the boundary.} for the geodesic orthogonal to the boundary, and $\delta$ is a positive constant.
Taking the limit $\delta \to 0$, the anomalous action (\ref{anb}) becomes
\begin{eqnarray}
S_{anom}[g] \mathop  \to \limits^{\delta  \to 0} \frac{1}{96\pi} \{ \int {{d^2}x\sqrt { - g} ( - \varphi \Box \varphi  - 2\varphi R) } + \int {{d^1}x\sqrt \gamma  (\varphi {n_\mu }{\nabla ^\mu }\varphi  - 4\varphi K)}  \}.\label{bac}
\end{eqnarray}
It turns out that we have exactly the same action as the previous one (\ref{SanomLocal})  expect for the additional boundary terms.
The boundary terms have no contribution on the stress tensor except on the boundary, and thus the obtained form of the stress tensor in $\cal M$ is the same as that without the boundary term.
Meanwhile, the boundary terms affect the boundary condition for the scalar field $\varphi$.
Equation (\ref{EOMofPhi}) can be rewritten in
\begin{align}
%L_{f_\delta} \varphi \equiv ( -\Box + \nabla^{\mu}f_\delta n_{\mu}n_{\nu}\nabla^{\nu} ) \varphi =
 -\Box \varphi + (n^{\mu} \nabla_{\mu} f_\delta)(n^{\nu} \nabla_{\nu} \varphi) + f_\delta \nabla_{\mu} n^{\mu} (n^{\nu} \nabla_{\nu} \varphi) %\nonumber \\
%= R + 2 \nabla_{\mu}(f_\delta Kn^\mu)
= R + 2 (n^{\mu} \nabla_{\mu} f_\delta)K + 2 f_\delta \nabla_{\mu} n^{\mu} K. \label{EOMofPhiDelta0}
\end{align}
Taking the limit $\delta \to 0$, we find the equations for $\varphi$
\begin{eqnarray}
\Box \varphi = -R,
\label{phiEq1}
\end{eqnarray}
with the boundary conditions\footnote{
These equations can be also obtained from the action (\ref{bac}) directly. Note that because $ - n^{\nu} \nabla_{\nu} f_\delta$ becomes Dirac delta function in the limit $\delta \to 0$, the terms proportion to it in the lhs and rhs of eq.(\ref{EOMofPhiDelta0}) should be balanced. }
\begin{eqnarray}
n^{\nu} \nabla_{\nu} \varphi = 2K, \qquad x \in  \Sigma.
\label{phiEq2}
\end{eqnarray}
This means that there is the additional boundary constraint on $\varphi$ which was not  taken into consideration in the previous works.

\section{Application to various spacetimes}\label{EX}
In this section, we apply our result to several common spacetimes, which are the flat, two-dimensional Schwarzschild, and de Sitter spacetimes. Since any two-dimensional spacetime can be described by the conformally-flat metric, we analyze the general conformally-flat spacetime at first.
Then, we see the application to the concrete spacetimes.

\subsection{General analysis}

Any metric of two-dimensional spacetime can be written in the conformal flat form:
%We consider a spacetime whose metric is described by the conformal form:
\begin{eqnarray}
d{s^2}  = F(t,r)( - d{t^2} + d{r^{2}}).\label{Mcf}
\end{eqnarray}
We consider the case in which the boundaries exist on $r=r_1$ and $r=r_2=r_1+L(>r_1)$.
The normal vector on the boundary is written in
\begin{eqnarray}
n^{\mu} =
\left(
\begin{array}{cc}
 0, & { F^{ -\frac{1}{2}} }
\end{array}
\right).
\end{eqnarray}
The Ricci scalar and extrinsic curvature on the boundary are, respectively,
\begin{eqnarray}
&&R = F^{-1} (-\partial^2_t \ln F + \partial^2_r \ln F),\\
&&K = \frac{1}{2} F^{-\frac{3}{2}} \partial_r F.
\end{eqnarray}

With the metric (\ref{Mcf}), eq.~(\ref{phiEq1}) can be rewritten as
\begin{eqnarray}
F^{-1} (-\partial^2_t \varphi + \partial^2_r \varphi)= F^{-1} (-\partial^2_t \ln F + \partial^2_r \ln F).
\end{eqnarray}
A particular solution of this equation is $\ln F(=: \varphi_p)  $, and thus the general solution for $\varphi$ is derived as
\begin{eqnarray}
&&\varphi = \varphi_p + \varphi_h, \label{nob}\\
&&\varphi_h:= A_1 r + A_2 t + A_3 + A_0 r t + \int^{\infty}_{-\infty} d\omega [ c_\pm(\omega) e^{i\omega t} e^{\pm i\omega r} ] + \int^{\infty}_{-\infty} d\omega [ d_\pm(\omega) e^{\omega t} e^{\pm \omega r} ] ,
\label{phiConfomllyFlat}
\end{eqnarray}
where $\varphi_h$ is the homogeneous solution satisfying $\Box \varphi_h = 0$. $A_0$, $A_1$, $A_2$, $A_3$ are real constants, $c_\pm(\omega)$ are constant functions satisfy $c_\pm(\omega) = c_\pm^*(-\omega)$, and $d_\pm(\omega)$ are real functions.

The boundary equation (\ref{phiEq2}) becomes
\begin{eqnarray}
F^{-\frac{1}{2}} \partial_r \varphi = F^{-\frac{3}{2}} \partial_r F.
\end{eqnarray}
With this boundary condition, the solution (\ref{nob}) is constrained as
\begin{eqnarray}
&&\varphi = \varphi_p + \varphi_0, \label{gbou}\\
&&\varphi_0:= A_2 t + A_3 + \sum\limits_{n =  - \infty }^\infty  {{c_n} \cos(\omega_n r)} e^{i\omega_n t},
\label{phiConfomllyFlatB}
\end{eqnarray}
where $\omega_n=\frac{{\pi n}}{L}$, $n\in\mathbb{N}$, $c_n$ are constants satisfy $c_n=c_{-n}^*$.
%by comparing (\ref{phiConfomllyFlat}) and (\ref{phiConfomllyFlatB}),
%we would find that the exponential modes and sinusoidal modes should be removed and modified individually.
%This difference is always there even we consider the boundary lay on spacial infinity.

The stress tensor of the trace anomaly (\ref{Tanom}) can be transformed as
\begin{eqnarray}
&&{\rm{T}}_{\mu \nu }^{anom}[\varphi=\varphi_p+\varphi_0;g_{\mu\nu}] = T_{\mu\nu}^{\varphi_p} + T_{\mu\nu}^{\varphi_0},\\
%&=& \frac{1}{{24\pi }}[ {g_{\mu \nu }}\Box\varphi + \frac{1}{4}{g_{\mu \nu }}({\nabla _\alpha }\varphi )({\nabla ^\alpha }\varphi ) - {\nabla _\mu }{\nabla _\nu }\varphi - \frac{1}{2}({\nabla _\mu }\varphi )({\nabla _\nu }\varphi ) ] \\
%&=& \frac{1}{{24\pi }}[ {g_{\mu \nu }}\Box\varphi_p + \frac{1}{4}{g_{\mu \nu }}({\nabla _\alpha }\varphi_p )({\nabla ^\alpha }\varphi_p ) - \frac{1}{2}({\nabla _\mu }\varphi_p )({\nabla _\nu }\varphi_p ) - {\nabla _\mu }{\nabla _\nu }\varphi_p \\
%&& + \frac{1}{4}{g_{\mu \nu }}({\nabla _\alpha }\varphi_0 )({\nabla ^\alpha }\varphi_0 ) - \frac{1}{2}({\nabla _\mu }\varphi_0 )({\nabla _\nu }\varphi_0 ) - {\nabla _\mu }{\nabla _\nu }\varphi_0 \\
%&& + \frac{1}{2}{g_{\mu \nu }}({\nabla _\alpha }\varphi_0 )({\nabla ^\alpha }\varphi_p ) - \frac{1}{2}({\nabla _\mu }\varphi_p )({\nabla _\nu }\varphi_0 ) - \frac{1}{2}({\nabla _\mu }\varphi_0 )({\nabla _\nu }\varphi_p )\\
&&T_{\mu\nu}^{\varphi_p}:= \frac{1}{{24\pi }}[ {g_{\mu \nu }}\Box\varphi_p + \frac{1}{4}{g_{\mu \nu }}({\nabla _\alpha }\varphi_p )({\nabla ^\alpha }\varphi_p ) - \frac{1}{2}({\nabla _\mu }\varphi_p )({\nabla _\nu }\varphi_p ) - {\nabla _\mu }{\nabla _\nu }\varphi_p, \\
&&T_{\mu\nu}^{\varphi_0}:= \frac{1}{4}{g_{\mu \nu }}({\nabla _\alpha }\varphi_0 )({\nabla ^\alpha }\varphi_0 ) - \frac{1}{2}({\nabla _\mu }\varphi_0 )({\nabla _\nu }\varphi_0 ) - {\partial _\mu }{\partial _\nu }\varphi_0 .
\end{eqnarray}
Note that there is no coupling term between $\varphi_p$ and $\varphi_0$, %(in $(t,r)$ frame),
i.e. $T_{\mu \nu}$ can be separated into $\varphi_p$ part and $\varphi_0$ part. As we will see later, $\varphi_p$ part indeed describes the vacuum polarization, while $\varphi_0$ part seems related to the excitations.
Since all $\varphi$'s in the stress tensor have at least one derivative,
$A_3$ does not affect the stress tensor.
Therefore, without loss of generality, hearinafter we set $A_3$ to be naught.
Furthermore, if we restrict $\varphi_0$ to be  $A_2 t$, the $\varphi_0$ part of stress tensor ($T_{\mu \nu}^{\varphi_0}$) would become stationary.
\footnote{$T_{\mu \nu}^{\varphi_p}$ might not be stationary in general because of time dependence of $F(t,r)$.}
This contribution is expected to be that of the thermal state.

\subsection{Examples}

Here, we derive the concrete values of the stress tensor in simple cases; the Minkowski, two-dimensional Schwarzschild and de Sitter spacetimes.
We take various boundary conditions and show that we can naturally get the stress tensor of various vacuum states.

\subsubsection{Minkowski (flat) spacetime}

\begin{figure}
\begin{minipage}{55mm}
\includegraphics[width=\linewidth,height=55mm]{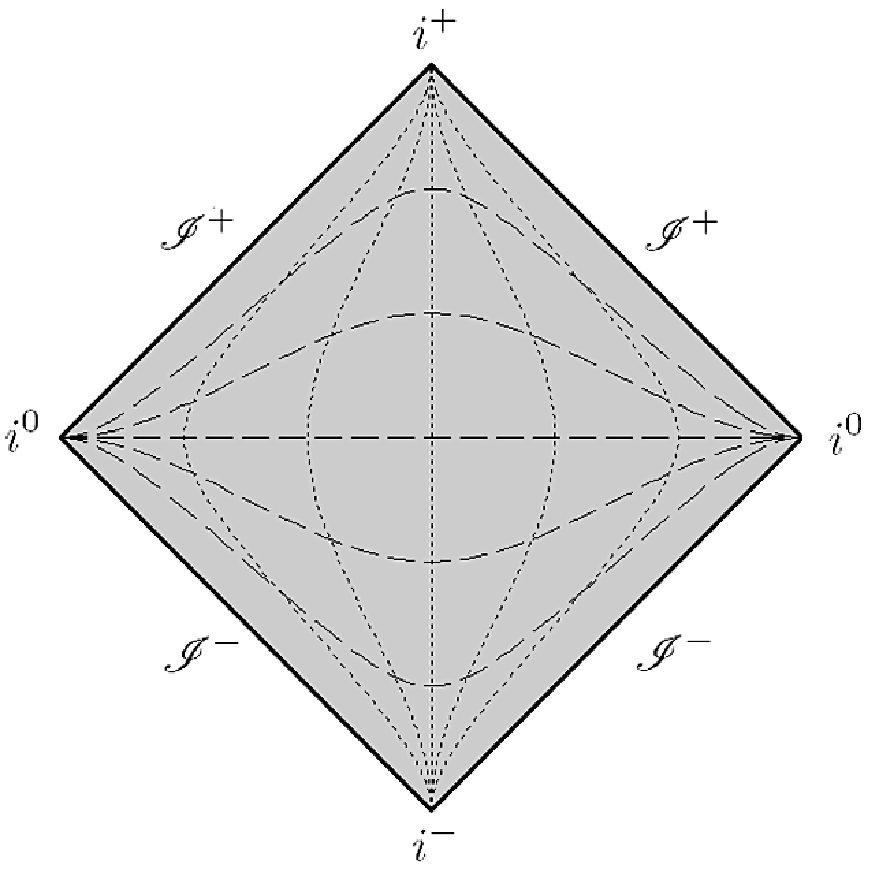}
\caption{Region corresponding to the Minkowski vacuum: The $(t,x)$ coordinate covers the whole Minkowski spacetime where $t =$ constant and $r =$ constant curves are drawn in dashed and dotted lines respectively. }
\label{fig1}
\end{minipage}
\begin{minipage}{55mm}
\includegraphics[width=\linewidth,height=55mm]{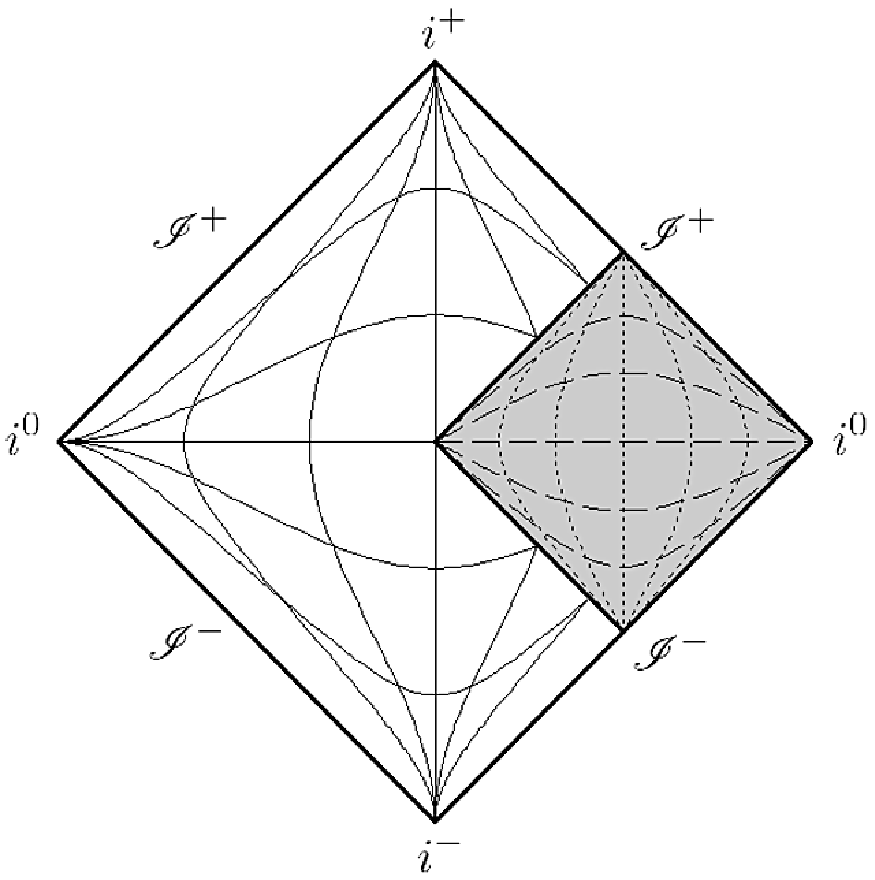}
\caption{Region corresponding to the Rindler vacuum: The $(T_R,R_R)$ coordinate covers only one quarter of Minkowski spacetime (Rindler wedge) where $T_R =$ constant and $R_R =$ constant curves are drawn in dashed and dotted lines respectively. }
\label{fig2}
\end{minipage}
\begin{minipage}{55mm}
\includegraphics[width=\linewidth,height=55mm]{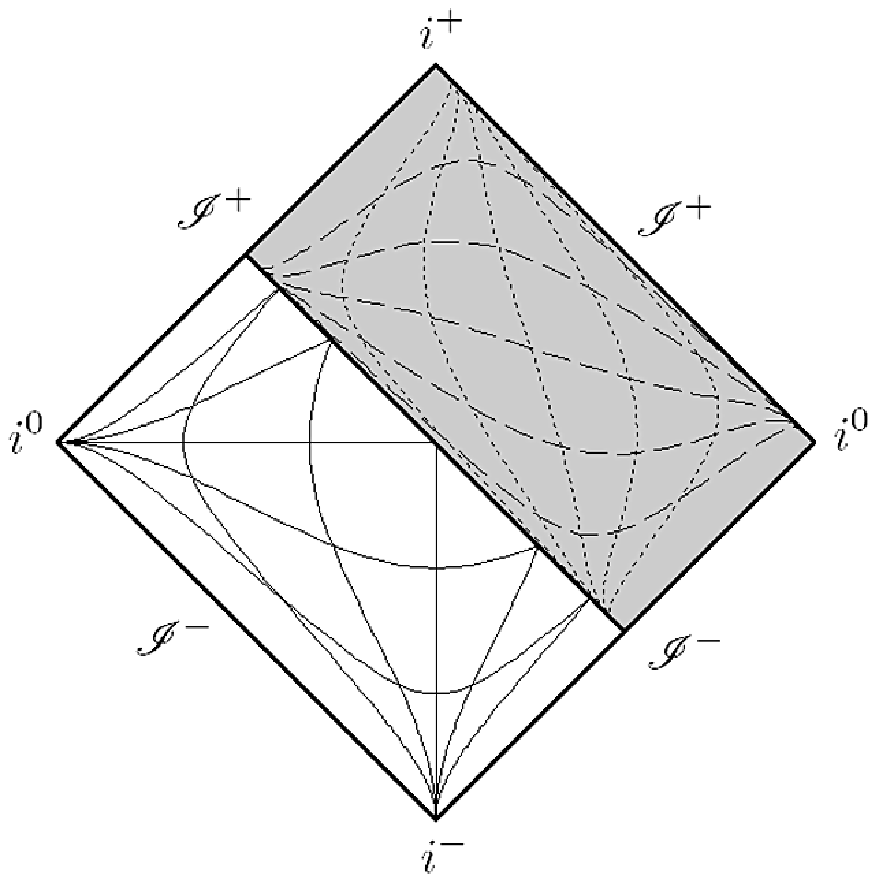}
\caption{Region corresponding to the Unruh-like vacuum: The $(T_U,R_U)$ coordinate covers a half of Minkowski spacetime where $T_U =$ constant and $R_U =$ constant curves are drawn in dashed and dotted lines respectively. }
\label{fig3}
\end{minipage}
\end{figure}

Minkowski spacetime is the simplest example. Let us consider it at first.
There are two famous vacua; the Minkowski vacuum (which based on the Cartesian coordinate) and the Rindler vacuum.
The vacuum based on the Cartesian coordinate is defined in the full region of Minkowski spacetime (see FIG.~\ref{fig1}), and thus we expect that the boundaries exist at two spatial infinities.
Meanwhile, the Rindler vacuum is defined in the Rindler wedge (see FIG.~\ref{fig2}).
One of boundary exists on the Rindler horizon and the other is at spatial infinity.
Moreover, for comparison with the discussion of the two-dimensional Schwarzschild spacetime that we will discuss later,
we consider another vacuum, which is the Unruh-like vacuum.
This is just the analog to the Unruh vacuun in the two-dimensional Schwarzschild spacetime;
one of the boundaries is the white hole horizon, and the other is spatial infinity.
The corresponding region is the sum set of the Rindler patch and the future Milne patch (see FIG.~\ref{fig3}).

To describe each region, we write the Minkowski metric in various forms:
\begin{eqnarray}
d{s^2} &=&  - d{t^2} + d{x^2} = - dU_f dV_f \label{Min}\\
 &=&  - \rho^2 du_fdv_f
 =  - \rho^2 ( - d{T_R}^2 + d{R_R}^2) \left(= - \rho^2 d{T_R}^2 + d{\rho^2}\right)\label{Rin}\\
 &=&  - V_f dU_fdv_f
 = V_f (-d{T_U}^2 + d{R_U}^2),\label{MinU}
\end{eqnarray}
where
\begin{eqnarray}
&&U_f := t - x, \qquad
V_f := t + x, \label{Mint}\\
&&\rho := (x^2-t^2)^{1/2}, \quad
u_f := - \log (-U_f), \quad
v_f := \log V_f, \quad
T_R := \frac{1}{2} (v_f+u_f), \quad
R_R := \frac{1}{2} (v_f-u_f),\quad \label{Rint} \\
&&T_U := \frac{1}{2} (v_f+U_f), \qquad
R_U := \frac{1}{2} (v_f-U_f). \label{MinUt}
\end{eqnarray}
The metric forms (\ref{Min}), (\ref{Rin}) and (\ref{MinU}) describe the regions of whole, Rindler patch and the sum set of the Rindler patch and the future Milne patch of Minkowski spacetime, and they are corresponding to the Minkowski vacuum, Rindler vacuum and Unruh-like vacuum, respectively.

$\ $

\paragraph{Minkowski Vacuum}

$\ $

The Minkowski vacuum is the lowest energy state defined in whole of Minkowski spacetime.
Therefore, we consider coordinate (\ref{Min}) with boundaries at $x=x_{\pm}$ and take the limit $x_\pm \to \pm \infty$.

From eq.(\ref{gbou}), the general solution can be written in
\begin{eqnarray}
\varphi = A_2 t + \int d\omega \, {{c(\omega)} \cos[\omega x]} e^{i\omega t}.
\label{phiMinkowskiB}
\end{eqnarray}
The stationary stress tensor can be obtained be setting $c(\omega)=0$ as
\begin{eqnarray}
T_{\mu\nu} &=&
\left(
\begin{array}{cc}
 -\frac{A_2^2}{4} & 0 \\
 0 & -\frac{A_2^2}{4} \\
\end{array}
\right) \qquad \mbox{in $(t,x)$ coordinate.}
\end{eqnarray}
As $A_2=0$, $T_{\mu\nu}$ becomes the same as that of the Minkowski vacuum, i.e. all components become zero.
Meanwhile, $A_2$ characterizes the temperature of the thermal equilibrium state.

$\ $

\paragraph{Rindler Vacuum}

$\ $

The Rindler patch is described by the metric (\ref{Rin}).
We consider the case where boundaries exist at $R_R= R_{\pm}$ and take the limit $R_\pm \to \pm \infty$.
Then the solution for $\varphi$ becomes
\begin{eqnarray}
\varphi = 2 R_R + A_2 T_R + \int  d\omega \, {{c(\omega)} \cos[\omega R_R]} e^{i\omega T_R}.
\end{eqnarray}
The stationary stress tensor (with respect to the Rindler time) is realized if $c(\omega)=0$,
and the corresponding stress tensor is:
\begin{eqnarray}
T_{\mu\nu} &=&
\left(
\begin{array}{cc}
 1-\frac{A_2^2}{4} & 0 \\
 0 & 1-\frac{A_2^2}{4} \\
\end{array}
\right) \qquad \mbox{in $(T_R,R_R)$ coordinate}, \\
&=&
\left(
\begin{array}{cc}
 -\frac{\left(A_2^2-4\right) \left(x^2+t^2\right)}{4 \left(x^2-t^2\right)^2} &
   \frac{\left(A_2^2-4\right) x t}{2 \left(x^2-t^2\right)^2} \\
 \frac{\left(A_2^2-4\right) x t}{2 \left(x^2-t^2\right)^2} & -\frac{\left(A_2^2-4\right)
   \left(x^2+t^2\right)}{4 \left(x^2-t^2\right)^2} \\
\end{array}
\right) \qquad
\mbox{in $(t,x)$ coordinate.}
\end{eqnarray}
For $A_2=0$, the result is the same as that corresponding to the Rindler vacuum state,
and $A_2$ characterizes the temperature of the ``thermal equilibrium state" based on the Rindler vacuum.
The condition $A_2=2$ gives the same result as that corresponding to the Minkowski vacuum state, and thus the vacuum of the Cartesian coordinate is a thermal state based on the Rindler vacuum.
This is consistent with the Unruh effect; the Rindler observer feels the thermal radiation in the Minkowski vacuum state.

%Rindler space and flat are indeed the same manifold(locally) (with different boundary). Therefore if you consider only eq.(\ref{phiEq1}), you will find no difference between these two cases. However, due to the difference of their boundary, the boundary constraints eq.(\ref{phiEq2}) for them are indeed different and will yield different vacuum solutions!\\

%Next we will consider another boundary in Minkowski space which has no explicit physical meaning but is interesting when we compare it with the similar boundary in Schwarzschild case.

$\ $

\paragraph{Unruh-like Vacuum}

$\ $

In the Schwarzchild black hole spacetime, we are sometimes interested in the vacuum state defined in the sum set of the outer region and the future trapped region, which gives the Unruh state.
To see the correspondence between the Minkowski spacetime and the two-dimensional Schwarzschild spacetime that we will discuss later, it is useful to consider the corresponding situation.
That is, we consider the sum set of the Rindler patch and the future Milne patch, which is described by the metric (\ref{MinU}).
The boundaries are set at $R_{U}=R_{\pm}$ and we take the limit $R_{\pm}\to\pm\infty$.
Then the solution for $\varphi$ becomes
\begin{eqnarray}
\varphi = \ln (t+x) + A_2 T_U  + \int d\omega \, {{c(\omega)} \cos[\omega R_U]} e^{i\omega T_U}.
\end{eqnarray}
The stress tensor of the thermal  state is expected to be obtained with the condition $c(\omega)=0$:
\begin{eqnarray}
T_{\mu\nu} &=&
\left(
\begin{array}{cc}
 \frac{1}{2}-\frac{A_2^2}{4} & \frac{1}{2} \\
 \frac{1}{2} & \frac{1}{2}-\frac{A_2^2}{4} \\
\end{array}
\right)
\qquad \mbox{in $(T_U,R_U)$ coordinate,}\\
&=&
\left(
\begin{array}{cc}
 \frac{4-A_2^2 \left((x+t)^2+1\right)}{8 (x+t)^2} & \frac{ 4 + A_2^2 \left((x+t)^2-1\right)}{8
   (x+t)^2} \\
 \frac{4 + A_2^2 \left((x+t)^2-1\right)}{8 (x+t)^2} & \frac{4-A_2^2 \left((x+t)^2+1\right)}{8
   (x+t)^2} \\
\end{array}
\right)\qquad
\mbox{in $(t,x)$ coordinate.}
\end{eqnarray}
The terms depending on $A_2$ appear in the diagonal part in $(T_U,R_U)$ coordinate, and it is traceless.
This implies that its energy flows along $\partial_{T_U}$, and thus, a thermal gas comoves along $\partial_{T_U}$.
The case with $A_2=0$ is expected to be the vacuum state of the region that we consider.
The stress tensor has the off-diagonal term in $(t,x)$ coordinate.
This means that we have energy flow in the vacuum state, which is corresponding to the Hawking radiation in the Unruh state of the black hole spacetimes.

\subsubsection{two-dimensional Schwarzschild spacetime}

The vacuum polarization in the black hole spacetime is one of the major interests in the quantum field theory on curved spacetimes.
As a simplified toy model, the two-dimensional Schwarzschild spacetime is often investigated,
where we consider the same metric as the time and radial components of the four-dimensional
Schwarzschild spacetime.
This geometry is not a solution of a gravity theory,\footnote{In two dimensional spacetime, general relativity is not well-defined.}
but it is fixed by hand.
The artificial spacetime is enough for the discussion of the renormalized stress tensor.
The causal structure in this two-dimensional Schwarzschild spacetime is the same as that in the four-dimensional Schwarzschild spacetime, and thus qualitatively we can expect that similar features of the vacuum polarization, such as  the Hawking radiation, appear.
Here, we study the vacuum polarization of the three familiar states; the Boulware, Hartle-Hawing and Unruh states.

\begin{figure}
\begin{minipage}{55mm}
\includegraphics[width=\linewidth,height=30mm]{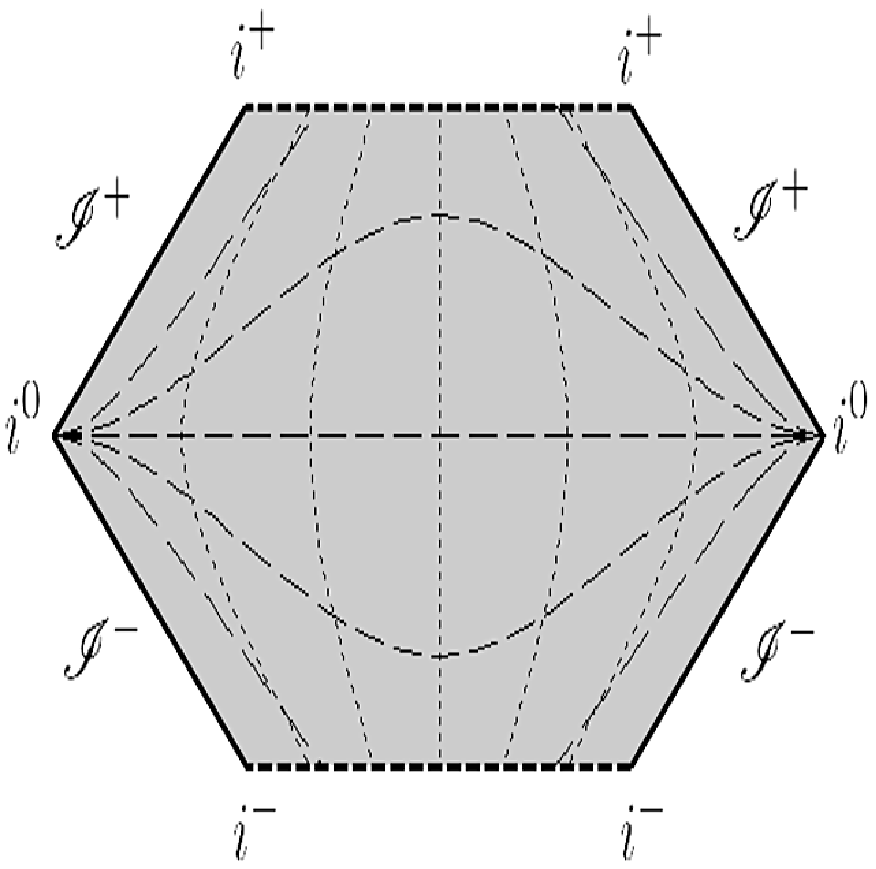}
\caption{Region corresponding to the Hartle-Hawking vacuum: The $(T_H,R_H)$ coordinate covers the whole two-dimentional Schwarzschild spacetime where $T_H =$ constant and $R_H =$ constant curves are drawn in dashed and dotted lines respectively. }
\label{fig4}
\end{minipage}
\begin{minipage}{55mm}
\includegraphics[width=\linewidth,height=30mm]{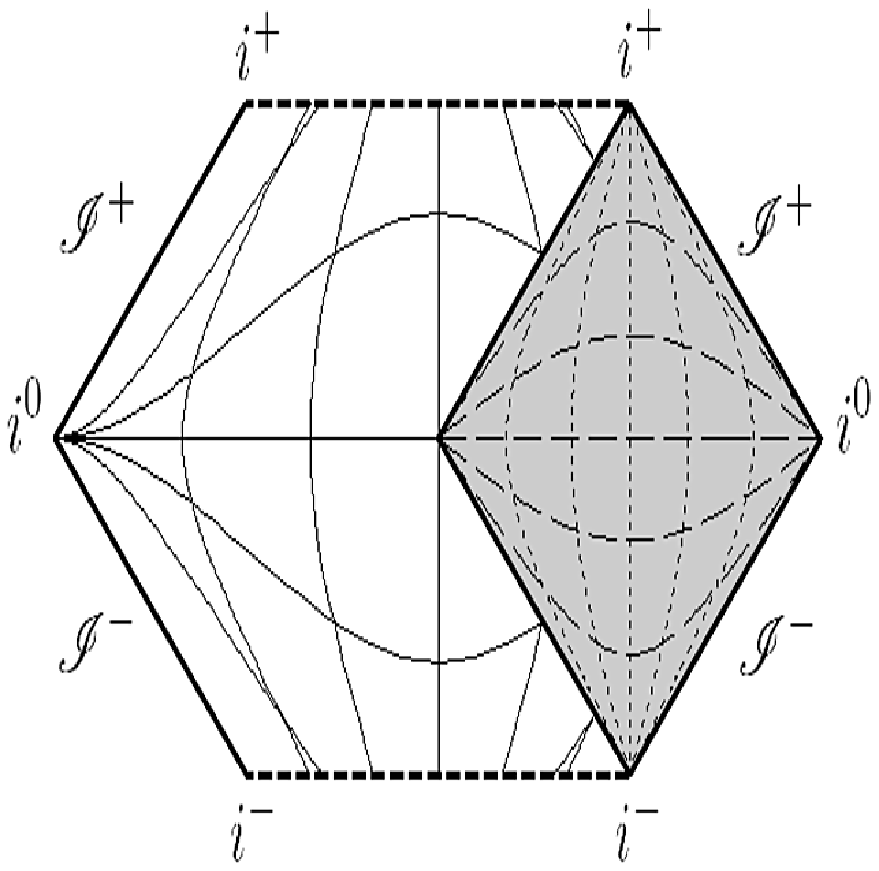}
\caption{ Region corresponding to the Boulware vacuum: The $(t,r)$ coordinate covers one quarter of two-dimentional Schwarzschild spacetime where $t =$ constant and $r =$ constant curves are drawn in dashed and dotted lines respectively. }
\label{fig5}
\end{minipage}
\begin{minipage}{55mm}
\includegraphics[width=\linewidth,height=30mm]{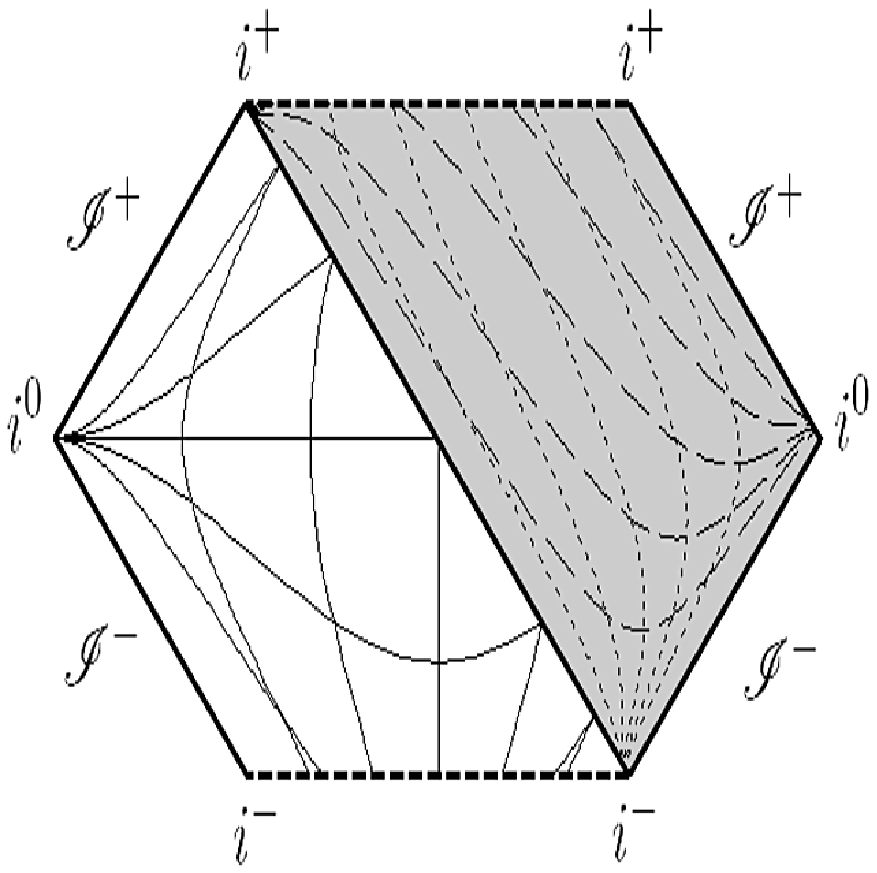}
\caption{ Region corresponding to the Unruh vacuum: The $(T_U,R_U)$ coordinate covers a half of two-dimentional Schwarzschild spacetime where $T_U =$ constant and $R_U =$ constant curves are drawn in dashed and dotted lines respectively. }
\label{fig6}
\end{minipage}
\end{figure}

In order to describe the corresponding regions to the three states, we write the two-dimensional Schwarzschild spacetime in various descriptions:
\begin{eqnarray}
d{s^2} &=&  - (1 - \frac{{2M}}{r})d{t^2} + {(1 - \frac{{2M}}{r})^{ - 1}}d{r^2}
 = (1 - \frac{{2M}}{r})( - d{t^2} + d{r^{*2}}) =  - (1 - \frac{{2M}}{r})dudv \label{BWc}\\
 &=&  - \frac{{32{M^3}}}{r}{e^{ - \frac{r}{{2M}}}}dUdV
 = - \frac{{32{M^3}}}{r}{e^{ - \frac{r}{{2M}}}}( - d{T_H}^2 + d{R_H}^2)\label{HHc}\\
 &=&  - \frac{{8{M^2}}}{r}{(\frac{r}{{2M}} - 1)^{\frac{1}{2}}}{e^{\frac{{t - r}}{{2M}}}}dUdv
 = \frac{{8M}}{r}{(\frac{r}{{2M}} - 1)^{\frac{1}{2}}}{e^{\frac{{t - r}}{{2M}}}}( - d{T_U}^2 + d{R_U}^2),\label{URc}
\end{eqnarray}
where
\begin{eqnarray}
&&r^{*} := r + 2M \ln (\frac{r}{2M} - 1), \qquad
u := t - r^{*}, \qquad
v := t + r^{*}, \label{BWt}\\
&&U := - {e^{\frac{-u}{4M}}}, \qquad
V := {e^{\frac{v}{4M}}}, \qquad
T_H := \frac{1}{2} (V+U), \qquad
R_H := \frac{1}{2} (V-U), \label{HHt} \\
&&T_U := \frac{1}{2} (v+U), \qquad
R_U := \frac{1}{2} (v-U). \label{URt}
\end{eqnarray}
The coordinates (\ref{BWc}), (\ref{HHc}) and (\ref{URc}) describe the outside of the black hole (see FIG.~\ref{fig5}), whole spacetime (see FIG.~\ref{fig4}) and the sum set of the outside and the future trapped region (see FIG.~\ref{fig6}), and
they are corresponding to the Boulware, Hartle-Hawking and Unruh states, respectively.
Comparing these coordinates (\ref{BWc}), (\ref{HHc}), (\ref{URc}) and the transformations  (\ref{BWt}), (\ref{HHt}), (\ref{URt}) with those of the Minkowski spacetime  (\ref{Min}), (\ref{Rin}), (\ref{MinU}), (\ref{Mint}), (\ref{Rint}) and (\ref{MinUt}), we can read the analog of the Boulware, Hartle-Hawing and Unruh vacua to the Rindler, Minkowski and Unruh-like vacua in the Minkowski spacetime, respectively.

$\ $

\paragraph{Hartle-Hawking Vacuum}

$\ $

The energy momentum tensor of the Hartle-Hawking state~\cite{Israel:1976ur,Hartle:1976tp} is defined in the whole spacetime, which is regular even at horizons and infinity, and thus state can be defined everywhere.
Therefore, the metric (\ref{HHc}) is the corresponding metric, which is regular everywhere.
We set the boundaries at $R_H=R_{\pm}$ and take the limit $R_{\pm}\to\pm\infty$.
Then the general solution (\ref{nob}) can be written in
\begin{eqnarray}
\varphi  = ln(1 - \frac{2M}{r}) - \frac{1}{2M} r^* + A_2 T_{H}  + \int d \omega \, {{c(\omega)} \cos[\omega R_H]} e^{i\omega T_H} .
\end{eqnarray}
Stationary stress tensor (in ($T_H,R_H$)-coordinate sense) is achieved if $c(\omega)$ vanishes, and it becomes
\begin{eqnarray}
T_{T_H T_H} &=& \frac{{64{M^4}}}{{{r^4}}}{{\rm{e}}^{ - \frac{r}{{2M}}}} - \left( {\frac{{48{M^4}}}{{{r^4}}} + \frac{{16{M^3}}}{{{r^3}}} + \frac{{4{M^2}}}{{{r^2}}}} \right){{\rm{e}}^{ - \frac{r}{M}}}\left( {{R_H}^2 + {T_H}^2} \right) - \frac{{{A_2}^2}}{4} , \\
T_{R_H R_H} &=& -\frac{{64{M^4}}}{{{r^4}}}{{\rm{e}}^{ - \frac{r}{{2M}}}} - \left( {\frac{{48{M^4}}}{{{r^4}}} + \frac{{16{M^3}}}{{{r^3}}} + \frac{{4{M^2}}}{{{r^2}}}} \right){{\rm{e}}^{ - \frac{r}{M}}}\left( {{R_H}^2 + {T_H}^2} \right) - \frac{{{A_2}^2}}{4} , \\
T_{T_H R_H} &=& T_{R_H T_H} =  \left( {\frac{{96{M^4}}}{{{r^4}}} + \frac{{32{M^3}}}{{{r^3}}} + \frac{{8{M^2}}}{{{r^2}}}} \right){{\rm{e}}^{ - \frac{r}{M}}}\left( {{T_H}{R_H}} \right) ,
\end{eqnarray}
\mbox{in $(T_H,R_H)$ coordinate,} and\\
\begin{eqnarray}
T_{t t} &=& - \left( {\frac{{7{M^2}}}{{{r^4}}} - \frac{{4M}}{{{r^3}}} + \frac{1}{{16{M^2}}}} \right) - \frac{{{A_2}^2}}{{64{M^2}}}\left( {{R_H}^2 + {T_H}^2} \right) , \\
T_{r r} &=& - {\left( {1 - \frac{{2M}}{r}} \right)^{ - 2}}\left( {\frac{1}{{16{M^2}}} - \frac{{{M^2}}}{{{r^4}}}} \right) - \frac{{{A_2}^2}}{{64{M^2}}}{\left( {1 - \frac{{2M}}{r}} \right)^{ - 2}}\left( {{R_H}^2 + {T_H}^2} \right) , \\
T_{t r} &=& T_{r t} =  - \frac{{{A_2}^2}}{{32{M^2}}}{\left( {1 - \frac{{2M}}{r}} \right)^{ - 1}}\left( {{T_H}{R_H}} \right) ,
\end{eqnarray}
\mbox{in $(t,r)$ coordinate.}

For $A_2=0$, the energy density is constant for the Killing observer (whose trajectory is tangent to $\partial_t$) outside the black hole, and the stress tensor is the same as that of the Hartle-Hawking vacuum state.
$A_2$ characterizes the thermal excitation based on the Hartle-Hawking vacuum.
%However, now no matter what value of $A_2$ you choose here, the result won't match the Boulware vacuum result, that means Boulware vacuum state is not the thermal equilibrium state in "Hartle-Hawking vacuum" sense.

$\ $

\paragraph{Boulware Vacuum}

$\ $

The Boulware vacuum~\cite{Boulware:1974dm} has the same asymptotic behavior as the Minkowski vacuum, while the stress tensor diverges on the horizon.
Thus, the state (and the quantum theory) is defined only outside the horizons.
The metric (\ref{BWc}) is the corresponding one.
We set the boundaries at $r^*=r^*_{\pm}$ and take the limit $r^*_\pm\to\pm\infty$.
The form of general solution (\ref{nob}) becomes
\begin{eqnarray}
\varphi  = \ln(1 - \frac{2M}{r}) + A_2 t + \int d\omega \, {{c(\omega)} \cos[\omega r^*]} e^{i\omega t} .
\end{eqnarray}
Imposing the stationary condition of the stress tensor, $c(\omega)$ should vanishes and  the stress tensor is derived as
\begin{eqnarray}
T_{\mu\nu} &=&
\left(
\begin{array}{cc}
 \frac{4 Mr-7 M^2}{r^4}-\frac{A_2^2}{4} & 0 \\
 0 & \frac{M^2}{r^4}-\frac{A_2^2}{4} \\
\end{array}
\right)\qquad
\mbox{in $(t,r^*)$ coordinate,} \\
&=&
\left(
\begin{array}{cc}
 \frac{4 Mr-7 M^2}{r^4}-\frac{A_2^2}{4} & 0 \\
 0 & \frac{M^2}{r^2 (r-2 M)^2}-\frac{A_2^2 r^2}{4 (r-2 M)^2} \\
\end{array}
\right) \qquad
\mbox{in $(t,r)$ coordinate.}
\end{eqnarray}
For $A_2=0$, the energy density has the minimum value, which corresponds to the Boulware vacuum state.
$A_2$ characterizes the temperature of the thermal equilibrium state based on the Boulware vacuum.
Similar to the relation between the Minkowski and Rindler vacua, for $A_2=\pm1/(2M)$, the resulting stress tensor is the same as that of the Hartle-Hawking vacuum state.
That is, Hartle-Hawking vacuum state is a thermal equilibrium state on Boulware vacuum.

$\ $

\paragraph{Unruh Vacuum}

$\ $

In the Unruh vacuum state~\cite{Unruh:1976db}, we take the Minkowski vacuum state at the past null infinity, while the stress tensor is regular on the black hole horizon but not on white hole horizon.
We can extend the state to the inside of the black hole but not of the white hole.
Therefore, the corresponding region is the sum of the outside of horizon and inside of black hole, which is described with the metric (\ref{URc}).
We set the boundaries at $R_U=R_{\pm}$ and take the limit $R_{\pm}\to\pm\infty$.
Then, the general solution is written in
\begin{eqnarray}
\varphi  = \ln(1 - \frac{2M}{r}) + \frac{1}{4M} (t-r^*) + A_2 T_U + \int d\omega \, {{c(\omega)} \cos[\omega R_U]} e^{i\omega T_U}.
\end{eqnarray}
The stationary stress tensor (in ($T_U$,$R_U$) sense) is obtained if $c(\omega)$ vanishes and it is written as
\begin{eqnarray}
T_{T_U T_U} &\!\!=\!\!&- \frac{{  M}}{{{r^4}}}\left( { { - r + \frac{3}{2}{M}}  - 16{M^2}{{\rm{e}}^{\frac{{t - r}}{{4M}}}}\sqrt {\frac{r}{{2M}} - 1}  + 2{M}{{\rm{e}}^{\frac{{t - r}}{{2M}}}}\left( {\frac{r}{{2M}} - 1} \right)\left( {{r^2} + 4Mr + 12{M^2}} \right)} \right) - \frac{{{A_2}^2}}{4}, \\
T_{R_U R_U} &\!\!=\!\!& -\frac{{  M}}{{{r^4}}}\left( { { - r + \frac{3}{2}{M}}  + 16{M^2}{{\rm{e}}^{\frac{{t - r}}{{4M}}}}\sqrt {\frac{r}{{2M}} - 1}  + 2{M}{{\rm{e}}^{\frac{{t - r}}{{2M}}}}\left( {\frac{r}{{2M}} - 1} \right)\left( {{r^2} + 4Mr + 12{M^2}} \right)} \right) - \frac{{{A_2}^2}}{4}, \\
T_{T_U R_U} &\!\!=\!\!& T_{R_U T_U} = - \frac{{  M}}{{2{r^4}}}\left( { - 2r + 3M - 2{{\rm{e}}^{\frac{{t - r}}{{2M}}}}\left( {r - 2M} \right)\left( {{r^2} + 4Mr + 12{M^2}} \right)} \right),
\end{eqnarray}
\mbox{in $(T_U,R_U)$ coordinate,} and\\
\begin{eqnarray}
T_{t t} &=& - \left( {\frac{1}{{32{M^2}}} + \frac{{7{M^2}}}{{{r^4}}} - \frac{{4M}}{{{r^3}}}} \right) - \frac{{{A_2}^2}}{8}\left( {1 + \frac{{r - 2M}}{{32{M^3}}}{{\rm{e}}^{\frac{{r - t}}{{2M}}}}} \right), \\
T_{r r} &=& - {\left( {1 - \frac{{2M}}{r}} \right)^{ - 2}}\left( {\frac{{ - {M^2}}}{{{r^4}}} + \frac{1}{{32{M^2}}}} \right) - \frac{{{A_2}^2}}{8}\left( {{{\left( {1 - \frac{{2M}}{r}} \right)}^{ - 2}} + {{\left( {1 - \frac{{2M}}{r}} \right)}^{ - 1}}\frac{r}{{32{M^3}}}{{\rm{e}}^{\frac{{r - t}}{{2M}}}}} \right), \\
T_{t r} &=& T_{r t} =  \frac{1}{{32{M^2}}}{\left( {1 - \frac{{2M}}{r}} \right)^{ - 1}} - \frac{{{A_2}^2}}{8}\left( {{{\left( {1 - \frac{{2M}}{r}} \right)}^{ - 1}} - \frac{r}{{32{M^3}}}{{\rm{e}}^{\frac{{r - t}}{{2M}}}}} \right),
\end{eqnarray}
\mbox{in $(t,r)$ coordinate.}

The lowest energy state with respect to $(T_U,R_U)$-coordinate is realized for $A_2=0$, and then the stress tensor is the same as that of Unruh vacuum state.
$A_2$ describes the thermal excitation for the Unruh observer (whose trajectory is tangent to $\partial_{T_U}$).

\subsubsection{de Sitter spacetime}

Here, we consider the stress tensor in de Sitter spacetime.
In cosmology, de Sitter spacetime approximately describes the beginning part of the Universe, i.e. inflation.
Meanwhile, de Sitter spacetime has the maximal symmetry, and thus has intriguing features. Therefore, de Sitter spacetime is interesting in both phenomenological and theoretical viewpoints.

In de Sitter spacetime, two vacua, the vacuum of the static chart and the Bunch-Davis vacuum, are often discussed.
We describe de Sitter spacetime with two different coordinates,
\begin{eqnarray}
d{s^2} &=&  - (1-H^2r_s^2)d{t_s^2} +(1-H^2r_s^2)^{-1} d{r_s^2} = (1-H^2r_s^2)(-d{t_s^2} + d{r_s^{2*}} ) \label{ST}\\
% &=& - d{t_f^2} + e^{2Ht_s} (1 - H^2 r_s^2) H^{-1} d{r_f^2}  \\
 &=& %e^{2H t_f} ( -d{\eta^2} + d{r_f^2} ) =
 - d{t_f^2} + e^{2H t_f} d{r_f^2} =
 \frac{1}{H^2 \eta^2} ( -d{\eta^2} + d{r_f^2} ) , \label{BD}
\end{eqnarray}
where
\begin{eqnarray}
&&r_s^{*} := \frac{\tanh^{-1}(Hr_s)}{H}, \\
&&r_f := r e^{-H t_f}, \qquad
\eta := - \frac{e^{-H t_f}}{H}, \qquad t_f := t_s + \frac{1}{{2H}}\log \left[ {{H^{ - 1}}\left( {1 - {H^2}{r_s^2}} \right)} \right] ,
\end{eqnarray}
and ``$s$" and ``$f$" mean the static and flat slicing charts, respectively.
The vacua with the coordinates (\ref{ST}) and (\ref{BD}) are corresponding to the vacuum of the static chart and the Bunch-Davis vacuum, respectively.

\begin{figure}
\begin{minipage}{55mm}
\includegraphics[width=\linewidth,height=55mm]{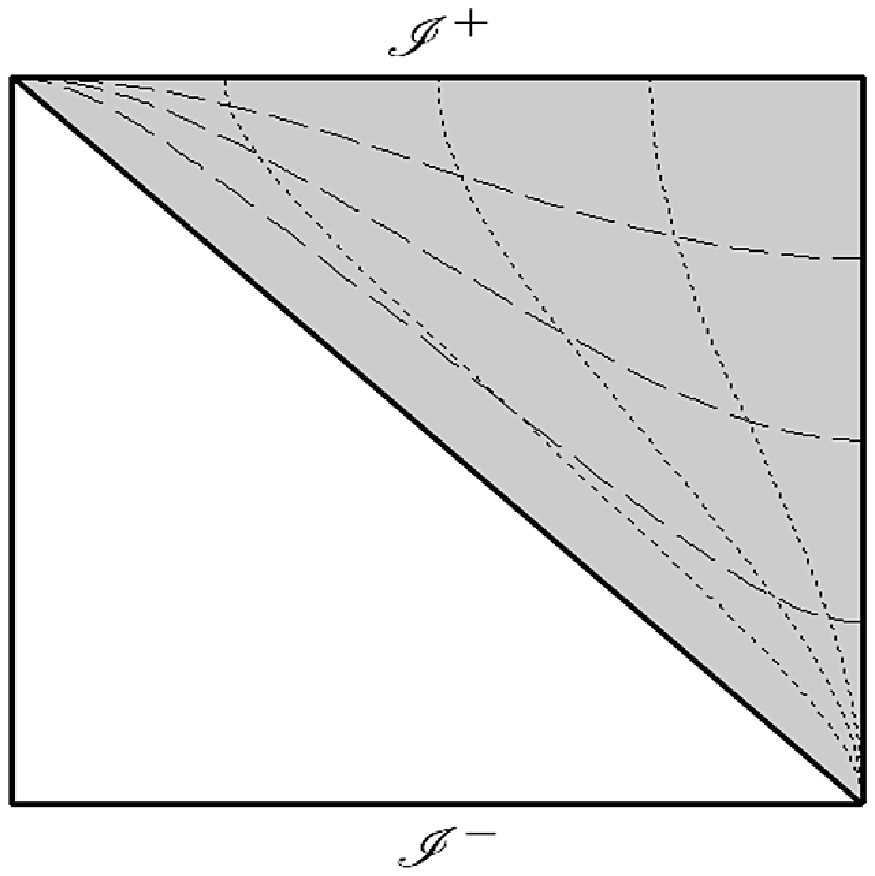}
\caption{Region corresponding to Bunch-Davies vacuum: The $(t_f,r_f)$ coordinate covers a half of de Sitter spacetime where $t_f =$ constant and $r_f =$ constant curves are drawn in dashed and dotted lines respectively. }
\label{fig7}
\end{minipage}
\hfil
\begin{minipage}{55mm}
\includegraphics[width=\linewidth,height=55mm]{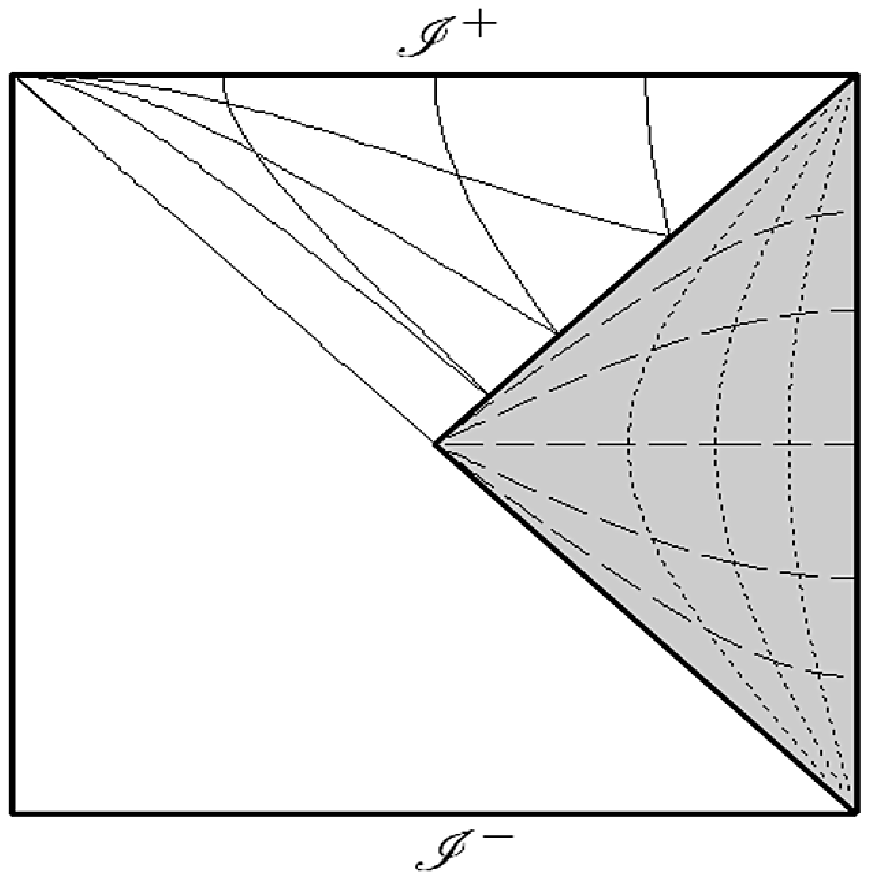}
\caption{Region of the static chart: The $(t_s,r_s)$ coordinate covers one quarter of de Sitter spacetime where $t_s =$ constant and $r_s =$ constant curves are drawn in dashed and dotted lines respectively. }
\label{fig8}
\end{minipage}
\end{figure}

$\ $

\paragraph{Bunch-Davis Vacuum}

$\ $

The vacuum state of the flat chart (\ref{BD}) is the so-called Bunch-Davis state~\cite{Bunch:1978yq}.
The flat chart (\ref{BD}) describes the region shown in FIG.~\ref{fig7}.
We set the boundaries at $r_f=r_{\pm}$ and take the limit $r_\pm\to\pm\infty$.
Then, the general solution (\ref{nob}) becomes
\begin{eqnarray}
\varphi = - 2 \ln(H\eta) + A_2 \eta + \int d\omega \, {{c(\omega)} \cos[\omega r_f]} e^{i\omega \eta}.
\end{eqnarray}
The stationary stress tensor with respect to the conformal time $\eta$ is obtained for $c(\omega)=0$ as
\begin{eqnarray}
T_{\mu\nu} &=&
\left(
\begin{array}{cc}
{H^2} - \frac{A_2^2}{4} e^{-2Ht_f} & 0 \\
 0 & - e^{2Ht_f} H^2 - \frac{A_2^2}{4} \\
\end{array}
\right)
\qquad \mbox{in $(t_f,r_f)$ coordinate,}\\
&=&
\left(
\begin{array}{cc}
 e^{2Ht_f} H^2 - \frac{A_2^2}{4} & 0 \\
 0 & - e^{2Ht_f} H^2 - \frac{A_2^2}{4} \\
\end{array}
\right)
=
\left(
\begin{array}{cc}
 \eta^{-2} - \frac{A_2^2}{4} & 0 \\
 0 & - \eta^{-2} - \frac{A_2^2}{4} \\
\end{array}
\right) \nonumber\\
&&\qquad\qquad\qquad\qquad\qquad\qquad\qquad\qquad\qquad\qquad\qquad\qquad\qquad
\mbox{in $(\eta,r_f)$ coordinate.}
\end{eqnarray}
The lowest energy state is realized for $A_2=0$, and then the stress tensor becomes the same as that of Bunch-Davis vacuum.
$A_2$ describes the thermal state with respect to the conformal time $\partial_\eta$.

$\ $

\paragraph{Vacuum of the static chart}

$\ $

The static chart (\ref{ST}) describes the region shown in FIG.~\ref{fig8}.
We set boundary at $r_s=r_{\pm}$ and take the limit $r_{\pm}=\pm\infty$.
Then the general solution (\ref{nob}) becomes
\begin{eqnarray}
\varphi =\ln( 1 - H^2 r_s^2 ) + A_2 t_s + \int d\omega \, {{c(\omega)} \cos[\omega r_s^*]} e^{i\omega t_s}.
\end{eqnarray}

The stationary stress tensor with respect to the Killing direction $\partial_{t_s}$ is obtained for $c(\omega)=0$, and it is derived as
\begin{eqnarray}
T_{\mu\nu} &=&
\left(
\begin{array}{cc}
2H^2 - H^4 r_s^2 - \frac{A_2^2}{4} & 0 \\
 0 & H^4 r_s^2 - \frac{A_2^2}{4} \\
\end{array}
\right) \qquad \mbox{in $(t_s,r_s^*)$ coordinate,} \\
&=&\left(
\begin{array}{cc}
2H^2 - H^4 r_s^2 - \frac{A_2^2}{4} & 0 \\
 0 & \frac{H^4 r_s^2 - A_2^2/4 }{( 1 - H^2 r_s^2)^2 }\\
\end{array}
\right) \qquad \mbox {in $(t_s,r_s)$ coordinate.}
\end{eqnarray}
Imposing $A_2=0$, the minimum energy state is realized and the stress tensor becomes the same as that of the vacuum state in static chart.
$A_2$ describes the thermal excitation on the static chart.
For $A_2=\pm2H$, the resulting stress tensor is the same as that of the Bunch-Davis vacuum state.
That is, Bunch-Davis vacuum state is a thermal equilibrium state based on static vacuum.

\section{Summary}\label{Sum}
In this paper,  we have derived the anomaly-induced action with the boundary effect by restoring the corresponding boundary terms to Lagrangian for the counter terms.
%
%After taking the boundary effect into consideration, we get the modified version of anomaly induced action which include some boundary terms now.
Although the boundary action seems not to revise the stress tensor in the region within boundary, there are indeed additional boundary constraints for the auxiliary field $\varphi$.
%It turns out that, by the boundary conditions derived from the boundary action, the general solution of $\varphi$ is constrained
Therefore, even though the functional form of the stress-tensor is the same as that without the boundary effect, due to the constraint of the argument $\varphi$, the stress-tensor is restricted.
This effect has not been noticed before, i.e. the degree of freedom in the general solution of $\varphi$ and the corresponding stress-tensor is actually much less than what people have considered before.

As examples, we have applied our result to several common spacetimes, flat, two-dimensional Schwarzchild, and de Sitter spacetimes, with various different boundaries.
In the previous works~\cite{Balbinot:1999vg,Mottola:2006ew}, although it was shown that the solution of auxiliary field can be tuned to describe the quantum vacuum state correctly in several examples, the principle for the correspondence behind was unclear.
We have shown that, in the spacetime that we have considered, the corresponding solution of the classical auxiliary field to the quantum state of the original field is the vacuum state with the proper boundary conditions.
This procedure is quite natural, because any tuning of solutions is not required anymore.

It is interesting to apply our result to some other topics, such as (dynamical) casimir effect.
Now since we know the correct relation between quantum states of the original field and the solution of the auxiliary field,
we can deal with the quantum effects on curved spacetime as the classical dynamics of the auxiliary field $\varphi$.
It can be expected that by using the classical anomaly-induced action, we can discuss the backreaction problem in semi-classical approaches without bothering the complicated calculation.
Another interesting direction is the formulation of the anomaly-induced action with boundary effect in four dimensional spacetime.
We expect that this approach in four-dimensional spacetime would be a powerful tool to investigate various physical-interested semiclassical problems, such as cosmology, semi-classical physics on black hole spacetime, and so on.
We leave these interesting explorations as future works.

\begin{acknowledgments}
The authors would like to thank Je-An Gu for useful comments.
K.~I. is supported by Taiwan National Science Council under Project No. NSC101-2811-M-002-103 and this work is also supported by Leung Center for Cosmology and Particle Astrophysics, Taiwan, R.O.C.
\end{acknowledgments}

%\end{fmffile}
\end{document}